\documentclass[reprint,aps,amsmath,amaaymb,prl,superscriptaddress,twocolumn]{revtex4-1}

\usepackage{amsmath}
\usepackage{mathrsfs}
\usepackage{bm}
\usepackage{graphicx}
\usepackage{subfigure}
\usepackage{color}
\begin{document}

\title{Self-energy effects and energy band theory for warm dense matter} 
\author{Chang Gao}
\affiliation{Center for Applied Physics and Technology, HEDPS, Peking University, Beijing 100871, China}
\affiliation{School of Physics, Peking University, Beijing 100871, China}

\author{Shen Zhang}
\affiliation{Center for Applied Physics and Technology, HEDPS, Peking University, Beijing 100871, China}
\affiliation{College of Engineering, Peking University, Beijing 100871, China}

\author{X. T. He}
\email{xthe@iapcm.ac.cn}
\affiliation{Center for Applied Physics and Technology, HEDPS, Peking University, Beijing 100871, China}
\affiliation{Institute of Applied Physics and Computational Mathematics, Beijing 100088, China}
\affiliation{IFSA Collaborative Innovation Center of MoE, Peking University, Beijing 100871, China}
\affiliation{Institute of Fusion Theory and Simulation, Zhejiang University, Hangzhou 310027, China}

\author{Wei Kang}
\email{weikang@pku.edu.cn}
\affiliation{Center for Applied Physics and Technology, HEDPS, Peking University, Beijing 100871, China}
\affiliation{College of Engineering, Peking University, Beijing 100871, China}
\affiliation{IFSA Collaborative Innovation Center of MoE, Peking University, Beijing 100871, China}

\author{Ping Zhang}
\email{zhang\_ping@iapcm.ac.cn}
\affiliation{Center for Applied Physics and Technology, HEDPS, Peking University, Beijing 100871, China}
\affiliation{Institute of Applied Physics and Computational Mathematics, Beijing 100088, China}
\affiliation{IFSA Collaborative Innovation Center of MoE, Peking University, Beijing 100871, China}

\author{Mohan Chen}
\affiliation{Center for Applied Physics and Technology, HEDPS, Peking University, Beijing 100871, China}
\affiliation{College of Engineering, Peking University, Beijing 100871, China}

\author{Cong Wang}
\affiliation{Center for Applied Physics and Technology, HEDPS, Peking University, Beijing 100871, China}
\affiliation{Institute of Applied Physics and Computational Mathematics, Beijing 100088, China}


\begin{abstract}
The energy band structures caused by self-energy shifting that results in bound energy levels broadening and merging in warm dense aluminum and beryllium are observed.
An energy band theory for warm dense matter (WDM) is proposed and a new code based on the energy band theory is developed to improve the traditional density functional method. 
Massive data of the equation of state and transport coefficients for WDM in medium and low Z have been simulated. 
The transition from fully degenerate to partially degenerate (related to WDM) and finally to non-degenerate state is investigated using the Lorenz number varying with the degeneracy parameter, and the lower and upper parameter boundaries for WDM are achieved. 
It is shown that the pressure ionization results in the Wiedemann-Franz law no longer available for WDM.
\end{abstract}

\maketitle

Warm dense matter (WDM)  \cite{Drake2006HEDP} is an important state in high energy density matter. 
In the past years, it has attracted great attention because of its importance in diverse physical systems, involving the interior of planets and some stars \cite{Guillot1999Science, Laio2000Science}, the near-isentropic implosion compression of a fusion-fuel capsule in inertial confinement fusion (ICF) \cite{Bodner1998PoP, Lindl1995PoP, He2016PoP}, and the intense-beam interaction with matters in laboratory\cite{Hoffmann2016MRE, Patel2003PRL, Ng2004PRL, Vinko2012Nature}. 
Depending on materials, WDM, in general, corresponds to density $\rho$ from normal solid density up to very high compression and temperature $T$ from a few electron voltage (eV) up to comparable to the Fermi temperature $T_F \propto \rho^{2/3}$. 
Here, WDM is defined in partially ionized state or partially degenerate state.
In WDM, both electrons (e) and ions (i) are screened by their opposite charges, formed the so-called quasiparticles, and their interactions undergo the influence of the screened Coulomb potential $V_{ei}^S$. 
Under such circumstances, the electron-ion coupling constant $\Gamma=ze^2/ak_BT$ and the degeneracy parameter $\vartheta=k_BT/\epsilon_F=T/T_F$ for WDM are orders of $\sim 1$, where $z$ is the ionized charge, $a$ is the averaged distance between electrons and ions, $k_B$ is the Boltzmann constant. 
$\epsilon_F=\epsilon_{F0}\eta^{2/3}=k_BT_F$ is the Fermi energy of the compressed matter with $\epsilon_{F0}=k_BT_{0F}$ being the Fermi energy in normal density $\rho_0$ and $\eta=\rho/\rho_0$ the ratio of the compressed density $\rho$ to $\rho_0$.
In such a system, the thermal de Broglie wavelength of electron $\Lambda_e=\frac{h}{\sqrt{2\pi m_ek_BT}}$ and the electron-ion screening length $r_s=\kappa_s^{-1}$ are both greater than the averaged electron distance $d_e=(\frac{4\pi}{3}n_e)^{-1/3}$, where $m_e$ is the electron mass, $n_e$ is the electron number density, $h$ is the Planck constant, $\kappa_s$ is the inverse screening length, and $r=|\vec{r_e}-\vec{r_l}|$. 
As a result, the quantum-many-body correlation effect for WDM must be considered. 
We also called such partially ionized WDM non-classical plasma (NCP).

The traditional first-principles quantum molecular dynamics (QMD) under the framework of the density functional theory (DFT) \cite{Hohenberg1964PR, Kohn1965PR, Mermin1965PR} at zero or finite temperatures has been developed to investigate properties of matter.
With the increase of density and temperature, however, a large number of bound-electron energy states in atoms are excited, and thousands of excited states appear in the vicinity of the ionization boundary (total energy $\varepsilon=0$). 
In the past long standing, it is almost impossible to investigate properties of WDM very well, invoking calculations of the equation of state (EOS) and transport coefficients in detail, by solving the Kohn-Sham (KS) equation to obtain each of the excited states. 
Usually, one has to employ orbital-free (OF) approximation, such as OFDFT code \cite{Lambert2006PRE, Recoules2009PRL, Wang2011PRL, karasiev2014CPC} involving Thomas-Fermi model, to deal with it. 
Thus, the shell-structure effect of the bound-energy levels disappears, resulting in loss of physical reality. 

\begin{table*}[th]
\caption{The self-energy shifting for warm dense Al varies with density $\rho$ (g/cc) and temperature $T$ (eV) and the bound-energy level shifting, broadening and merging are shown to compare with the isolated-atom model.}
\begin{ruledtabular}
\begin{tabular}{cccccccccccc}
& \multicolumn{2}{c}{states} &\multicolumn{9}{c}{energy levels (eV)} \\
\hline
& $\rho$ (g/cm$^3$) & $T$ (eV) & 1s & shifting & broadening & 2s & shifting & broadening & 2p & shifting & broadening \\
& \multicolumn{2}{c}{isolated atom} & -1606 & & & -123 & & & -88 & & \\
& 10.0 & 6.71 & -1460 & 146 & 10 & -68 & 55 & 16 & -30 & 58 & 21 \\
& 50.0 & 19.62 & -1374 & 232 & 17 & 7 & 130 & 126$^*$ & 59 & 147 & 126$^*$ \\
& 100.0 & 31.14 & -1298 & 308 & 28 & 50 & 173 & 218$^*$ & 141 & 229 & 218$^*$ \\
\end{tabular}
\end{ruledtabular}
\label{table: energy level}
\end{table*}

In this Letter, we show that the self energy \cite{Kremp2006Book}, an interaction energy of a test particle in NCP causes significant broadening and shifting of bound-energy levels toward the ionization boundary with the increase of density and temperature. 
We found that these levels in high-excited states are merged into broader energy-band structures, where the detailed configuration in atom disappears, while inner-shell energy bands shifted and broadened are still degenerate and discrete, resulting in the so-called partial degeneracy of WDM. 
Effects of self-energy shifting and bound-energy merging for WDM have been observed in experiments \cite{Hall1998EPL, Zhao2013PRL, DaSilva1989PRL, Nantel1998PRL}.
We proposed an energy band theory for WDM and developed a new code, which involves the Bloch wave, the scattering wave that is pivotal for transport behavior, and the plane wave, to improve previous extend first-principles molecular dynamics (ext-FPMD) method. 
We have achieved massive data of EOS and electrical and thermal conductivities for WDM in medium and low Z with great reduction of computing time and investigated the transport property of WDM using the Lorenz number varying with the degeneracy parameter $\vartheta$. 
It is found that the Wiedemann-Franz (W-F) law \cite{WF1853} is no longer available for WDM. 
From the $L$ number, we for the first time achieved lower and upper parameter boundaries of WDM, which are very important to select effective models and exactly describe properties of WDM. 

\begin{figure}[b]
\includegraphics[width=0.45\textwidth]{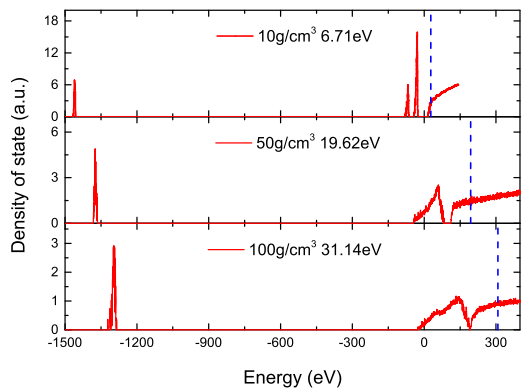}
\caption{Density of state (DOS) vs total energy $\varepsilon$~(eV) for warm dense Al varies with density and temperature, where the blue dished lines correspond to the chemical potential of Al.}
\label{fig: energy level}
\end{figure}

To identify the energy band structure, we concretely investigate the effect of the self-energy shift on the bound energy levels obtained from the solution of the Kohn-Sham equation of the thermal DFT \cite{Giannozzi2009QE} for warm dense aluminium (Al). 
It is shown in the Table~\ref{table: energy level} and Fig.~\ref{fig: energy level} for three sets of density $\rho$ and temperature $T$, where the change of the $n$-th bound energy level $\varepsilon_n$ caused by pressure ionization is compared with the isolated atomic model. 
In case (a) of Table~\ref{table: energy level}, where $T=6.71$~eV and $\rho=10$~g/cc, the 1s bound energy level is $\varepsilon_{1s}=-1460~\text{eV}~(-1606~\text{eV})$, here and below the data in the bracket are for the isolated atom model, correspondingly, the self-energy shift $\delta\varepsilon_{1s}=146~\text{eV}$ and the level broadening $\Delta_{1s}=10~\text{eV}$, as seen in Table~\ref{table: energy level}. 
For higher levels of 2s and 2p, bound energies are $\varepsilon_{2s}=-68~\text{eV}~(-123~\text{eV})$ and $\varepsilon_{2p}=-30~\text{eV}~(-88~\text{eV})$. 
Correspondingly, the self-energy shift appears in $\delta_{2s}=55~\text{eV}$ and $\delta_{2p}=58~\text{eV}$ and the level broadening is $\Delta_{2s}=16~\text{eV}$ and $\Delta_{2p}=21~\text{eV}$. 
It is shown that the bound levels moving clearly toward the ionization boundary due to the self-energy shifting result in the ionization energy lowering or pressure ionization. 
Here and below, we have neglected the electron spin-orbit coupling effect that causes the energy level split smaller. 
As for 3s3p levels and the highly-excited $n>3$, they have been completely ionized states and merged into a continuous plane wave distribution as $\varepsilon>\varepsilon_c$, where $\varepsilon_c$ greater than the Coulomb potential energy is an energy inflexion to separate the plane wave from 2s2p bands, as seen in Fig.~\ref{fig: energy level}(a). 
However, in the isolated atomic model 3s3p levels are still split, see Table~\ref{table: energy level}.
With further increasing density and temperature as seen in the case (b) of Table~\ref{table: energy level}, where $\rho=50$~g/cc and $T=19.62$~eV, the pressure ionization effect is quite significant. 
The 1s level further broadens and moves toward the ionization boundary but it is still discrete. 
The 2s2p levels have merged into an energy band and most of them manifest the feature of the scattering wave that interacts with the Coulomb potential in the energy interval $\varepsilon=0~\textrm{-}~\varepsilon_c$. 
While the 3s3p spectrum in $\varepsilon>\varepsilon_c$ is still the plane wave . 
As a result, two energy bands with a gap of tens eV in the both sides of the ionized boundary are formed, as shown in Fig.~\ref{fig: energy level}(b). 
In the case (c) of Table~\ref{table: energy level}, where $\rho=100$~g/cc and $T=31.14$~eV, the 2s2p band is completely ionized and two bands of 3s3p and 2s2p connect in their bottom, as seen in Fig.~\ref{fig: energy level}(c). 

We further confirmed from the above results that firstly, the self energy shifting causes the bound energy level broadening and shifting toward the ionization boundary with a rise of density and temperature, resulting in the bound energy lowering or the pressure ionization that mainly depends on the increase of density as clearly seen in Table~\ref{table: energy level} compared with levels of the isolated atom. 
Secondly, the bound energy level overlapping due to self-energy broadening and energy level merging results in the energy band structures. 
Thirdly, the gap between 2s2p band and conduction band is as high as tens of eV, but decreases with the increase of density and temperature. 
Thus, most of ionized electrons with large kinetic energy in the 2s2p band may directly transit to the conduction band 3s3p, similar to semiconductor. 
Therefore, in fact, the system of WDM exists in three kind of energy bands: the bound energy bands ($\varepsilon<0$), the scattering wave band ($\varepsilon$ in $0~\textrm{-}~\varepsilon_c$) and the band ($\varepsilon>\varepsilon_c$) related to the plane wave. 
Energy band theory can greatly save the computing time to compare with traditional DFT that thousands of high excited energy levels are unable to deal with in detail, and is beneficial for the study of properties of WDM.

\begin{figure}[t]
\includegraphics[width=0.5\textwidth]{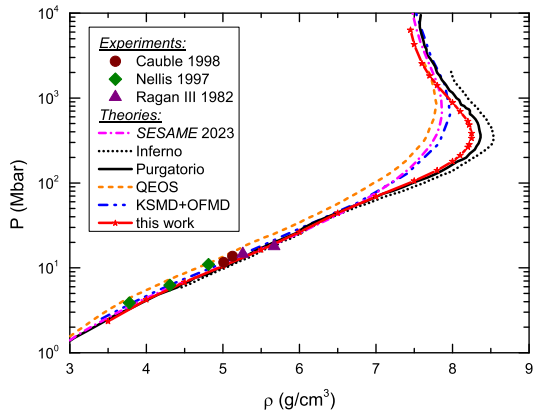}
\caption{Calculated $p$-$\rho$ Hugoniot curve for warm dense Be. The calculated results by ext-FPMD (red stars), laser experiment by Cauble et al. (wine points), explosive experiments by Nellis et al. (olive diamonds) and by Ragan III (purple upper triangles) are shown. Other models by \textit{SESAME} 2023 (magenta dashed-dotted line), QEOS (orange dashed line), Inferno (black dotte line), Purgatorio (black solid line) and KSMD+OFMD (blue dashed line) are also compared.}
\label{fig:Be-hugo}
\end{figure}

Based on the energy band structures, we now deal with the KS equation under thermal DFT framework \cite{Frontiers2014} in the form
\begin{small}
\begin{equation}
\left[-\frac{1}{2}\nabla^2+V^{\text{h}}[n]+V^{\text{xc}}[n]+V^{\text{ei}}(r)\right]\psi_i(r)=\epsilon_i\psi_i(r),
\end{equation}
\end{small}where $V^{\text{h}}[n]$ is Hartree energy, $V^{\text{xc}}$[n] is exchange-correlation energy, $V^{\text{ei}}(r)$ is the screened electron-ion interaction potential, $\epsilon_i$ is the eigenvalue, $\psi_i(r)$ is the electronic wave function, and $n$ is the electronic density. 
The solution for the KS equation is
\begin{equation}
\psi(\epsilon,r)\sim u(r)+\psi_{sc}(\epsilon,r)+\psi(\epsilon,r)_{bb},
\end{equation}
where $\psi(\epsilon,r)_{bb}$ is the bound-state wave function in energy $\varepsilon<0$ and is expressed by the Bloch wave as in traditional energy band theory of condensed matter physics, $\psi_{sc}(\epsilon,r)$ is the scattering wave scattered on the Coulomb potential in energy interval of $0~\textrm{-}~\varepsilon_c$ and $u(r)\sim \text{exp}(i\bm{k}\bm{r})$ is the plane wave in energy $\varepsilon>\varepsilon_c$. 

In a previous work, we developed a code called the ext-FPMD \cite{Zhang2016PoP}, in which only the Bloch waves and the plane waves under the Born approximation are considered by solving the KS equation of thermal DFT. 
The plane wave approximation is quite effective for calculating EOS of WDM with the increase of density and temperature because a lot of free-motion electrons with the particle number far greater than ions and the high kinetic energy provide main contributions to pressure and inner energy.
By using ext-FPMD, we have accomplished a large number of simulations and obtained massive data of EOS agreed with experiments very well for WDM in medium and low Z \cite{Zhang2016PoP, Gao2016PRB}, where the effect of the shell structure is clearly observed while it disappears using OFDFT. 
It has been shown that in most cases, calculations of EOS by ext-FPMD can greatly reduce computing time while the errors are within $\sim$ 1-2\% as $T \ge T_F$ and density compression $\eta=\rho/\rho_0\gg1$. 
On the other hand, the pressure ionization generates a lot of ionized electrons and the transport effect is very important to understand transport properties of WDM, in which the scattering wave is pivotal. 


Recently, we developed a new code, called energy band theory program (EBTP), based on energy band theory to solve the Kohn-Sham equation of thermal DFT for WDM. 
In this code we involve the scattering wave in the energy interval $\Delta\varepsilon=0$ - $\varepsilon_c$ besides the Bloch wave in $\varepsilon<0$ and the plane wave in $\varepsilon\ge\varepsilon_c$ 
Here, the Kubo-Greenwood formula (KGF) for optical description is applied to deal with the scattering wave, details will appear elsewhere. 

In the present simulation of EOS by EBTP, we obtained the $p$-$\rho$ Hugoniot curve for warm dense beryllium (Be) in Fig.~\ref{fig:Be-hugo}, which is an important material for the ablator of the fuel capsule in ICF. 
Our calculated results agreed with experiments very well by Ragan \cite{Ragan1982PRA}, Nellis \cite{Nellis1997JAP}, and Cauble \cite{Cauble1998PRL}, and as well with simulations in the range of pressure $< 150$~Mbar by Purgatorio \cite{Wilson2006JQSRT} and Inferno \cite{Liberman1979PRB}. 
The curve of Be possesses only one turning point, where the pressure $P\approx350~\text{Mbar}$ and about 50\% of 2 electrons in the K shell have been ionized. 
However, the curve, like most experiments, was performed by the shock compression only once. 
This shock is so strong that at the turning point the temperature reaches $T=90$~eV~$>$~$T_F\approx40$~eV while the maximal density compressed ratio is only $\eta=4.5$ with the degeneracy parameter $\vartheta\approx2.25$. 
It has approached the boundary of the nondegenerate state as discussed later in this text. 
Therefore, the compression by shock once produces WDM only in a narrow range of density and temperature. 
However, in most important cases, such as in the implosion compression of ICF and in the isentropic compression in laboratory, WDM is of density from the normal density to over 500~g/cc and temperature $T$ from a few eV to Fermi temperature $T_F\sim\rho^{2/3}\sim$keV \cite{Bodner1998PoP, Lindl1995PoP, He2016PoP}.
Thus, it has a wider range of density and temperature, where the upper boundary of WDM can reach $\vartheta\gg 1$, 1.e., $T\gg T_F$ far greater than that in the shock compression only once. 
Our simulations showed that it is almost impossible to investigate EOS of WDM, in general, as $\vartheta$ begins in $\vartheta\le\sim1$ without the plane wave approximation. 

\begin{figure}[t]
\includegraphics[width=0.45\textwidth]{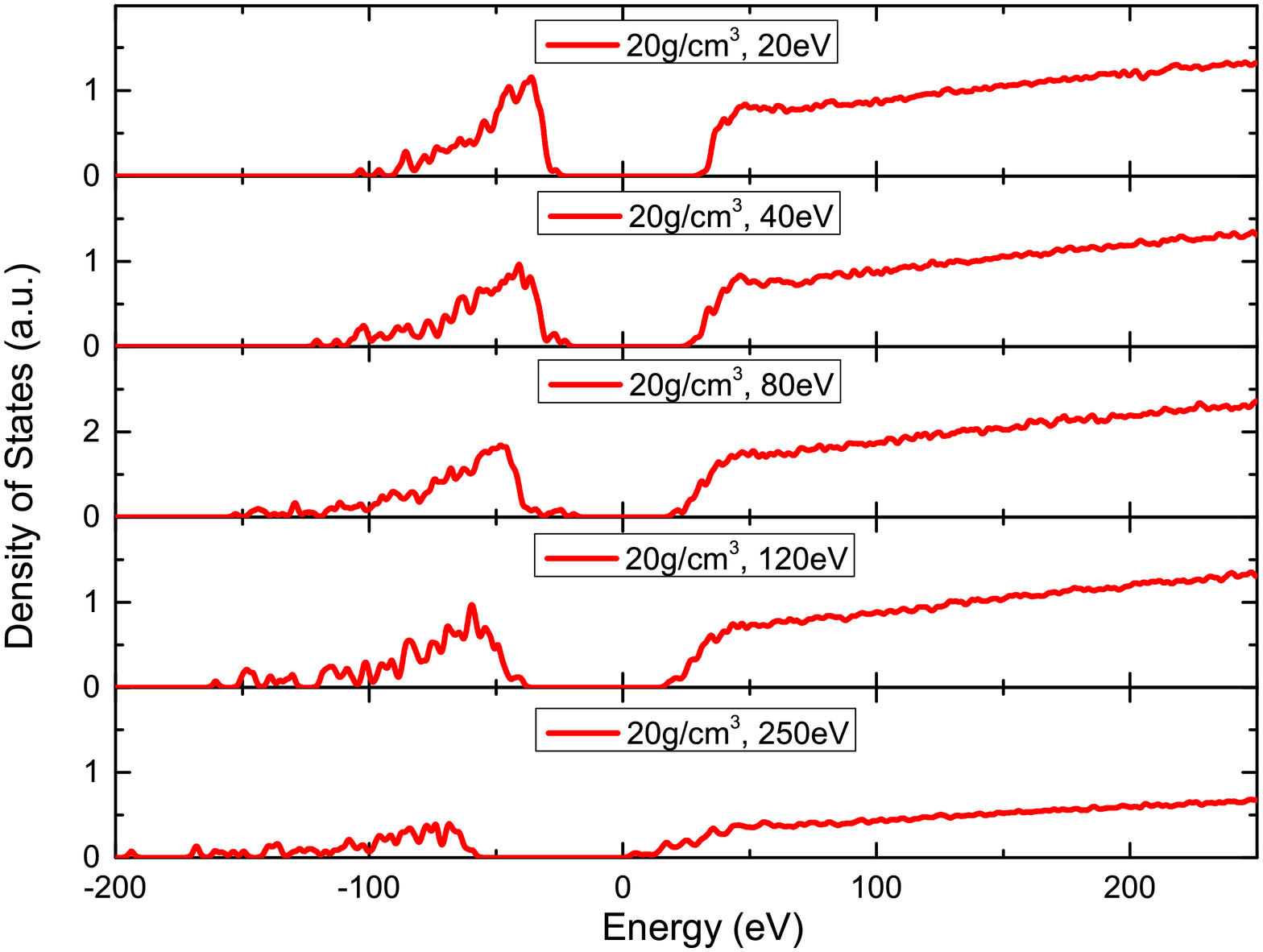}
\caption{Density of state (DOS) vs total energy $\varepsilon$~(eV) for warm dense Be varies with temperature.}
\label{fig:Be-dos}
\end{figure}

\begin{figure}[t]
\includegraphics[width=0.5\textwidth]{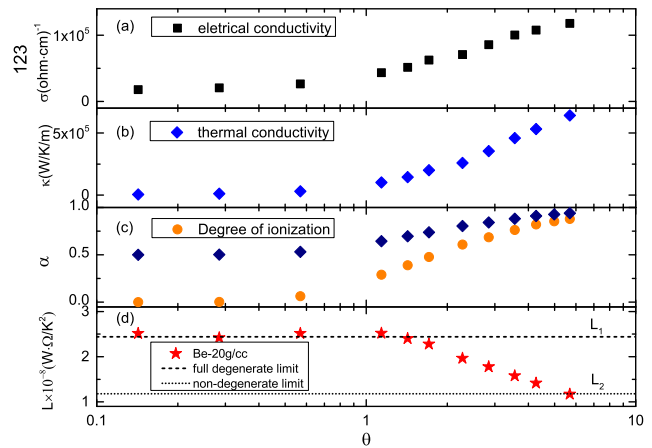}
\caption{Calculated electrical conductivity $\sigma$, thermal conductivity $\kappa$, degree of ionization $\alpha$, and Lorenz number $L$ of Be at density $\rho=20$ g/cc vs $\theta$. (a) $\sigma$, (b) $\kappa$, (c) $\alpha$, and (d) $L$. The dashed line in (d) is for the W-F law in the fully degenerate state, and the dotted line in (d) is for the W-F law in the non-degenerate state.}
\label{fig: Lorenz number}
\end{figure}

Using the EBTP approach, we now investigate transport properties of warm dense Be which has 2 conduction electrons in 2s and the 1s bound energy is of $\sim$111 eV (for isolated atom) in normal density $\rho_0=1.84$~g/cc. 
We now simulate its thermal conductivity $\kappa$ and electrical conductivity $\sigma$ under the condition of the compressed density $\rho=20$~g/cc with Fermi energy $\epsilon_F=72.2$~eV. 
In the present case, the pressure ionization, where the degree of ionization $\alpha$ is defined as the ionized fraction of the 1s electrons, results in significant effect on $\sigma$ and $\kappa$. 
The results are plotted in Fig.~\ref{fig: Lorenz number}, where Fig.~\ref{fig: Lorenz number}(a) and Fig.~\ref{fig: Lorenz number}(b) are for electrical conductivity $\sigma$ (black diamonds) and electronic thermal conductivity $\kappa$ (blue prismatic) respectively, and Fig.~\ref{fig: Lorenz number}(c) is for the degree of ionization $\alpha$ (red solid circle). 
We observed from Fig.~\ref{fig:Be-dos} that the pressure ionization results in the energy band structure for 1s energy levels and the peak of this energy band has moved from initial -111 to -49~eV at $\vartheta=1.14$ ($T=80$~eV) in DOS due to self energy shift, where a few of 1s electrons begin to transit to the scattering wave energy band as $\vartheta=1.71$ ($T>\sim 120$~eV) and then $\sim 75$\% of 1s electrons become the scattering wave at $\vartheta=3.6$. 
We also observed that the behavior of the electrical conductivity and thermal conductivity is quite different in both sides of $\vartheta\sim 0.65$.
In the region of $0<\vartheta \le \sim 0.65$ ($T=0.65T_F=45.6$~eV), $\alpha$ is almost 0 and is $\sim$ 0.063 at $\vartheta\sim 0.65$ in which pressure ionization just began, and $\sigma$ and $\kappa$ are small, where only 2 conduction electrons contribute to them. 
While in the region of $\sim 0.65<\vartheta\le \sim 5.8$ which corresponds to the degree of ionization $\alpha \sim 0.9$, as seen in Fig.~\ref{fig: Lorenz number}(c), $\sigma$ and $\kappa$ increase monotonously with the increase of $\alpha$ and are close to saturation at $\alpha=\sim 0.9$. 
As mentioned in previous discussion in calculations of EOS, we also encountered difficulties in the present calculation using the traditional method as $\vartheta$ begins in $\le\sim 1.0$ due to numerically intractable treatment \cite{Wang2011PRL}. 
While the new code EBTP results in significantly time-save with high precision in about 1\% compared with the traditional thermal DFT. 

In the following, with calculated $\sigma$ and $\kappa$ we investigate behavior of the Lorenz number $L$ in the whole range of the degeneracy parameter $\vartheta$. 
The $L$ number connects $\kappa$ with $\sigma$ and is written in the form of $L=\frac{K}{\sigma T}=\gamma(\frac{k_B}{e})^2$ \cite{Lorenz1872, Kumar1993JMS}, where $\gamma$ is the function of $\vartheta$. 
We also discuss features of the W-F law based on the study of the $L$ number and try to acquire the degeneracy parameter boundaries of WDM in high energy density matter.

The $L$ number has been widely applied to understand transport properties of materials, such as the carrier transport behavior in non-degenerate and degenerate semiconductors or metals \cite{Golinelli1996PRL, Thesberg2017PRB} and the heat flux in Earth's core \cite{Pozzo2012Nature, Konopkova2016Nature}, which is very important  to understand the Earth evolution etc. 
Here we use it to explore transport properties of WDM.

It is well known that, in the cases of the fully degenerate and non-degenerate, the $L$ number is unchanged with $\vartheta$ and the W-F law is available. 
Usually, using the W-F law or the L number one can obtain $\kappa$ from $\sigma$ that is easily measured. 
In previous works \cite{Wang2011PRL,Wang2013PREa,Wang2013PREb}, we have simulated the transport properties of WDM using thermal DFT, however, it is obliged to use the OFDFT as $\vartheta>\sim1.0$. 
We now have the ability to understand the transport nature of WDM in $\vartheta > \sim1.0$ with significant time saving by the new code EBTP. 
Before discussing the simulation results, we will first investigate the $L$ number characteristic analytically from physical view. 
For convenience, we divide the degeneracy parameter $\vartheta$ into three regions. 
In the region of $\vartheta=0~\textrm{-}~\vartheta_{lb}$, the system is in the full degenerate state and is governed by the quantum effect, where $\vartheta_{lb}$ is the lower boundary of the partial degeneracy. 
According to the W-F law, $\gamma$ is a constant of $\frac{\pi^2}{3}$ and the Lorenz number $L_1=2.44 \times 10^{-8}$(W$\cdot\Omega$)/K$^2$ is unchanged with $\vartheta$ \cite{WFlaw1961}. 
In the region of $\vartheta>\vartheta_{ub}$, where $\vartheta_{ub}$ is the upper boundary of the partial degeneracy, the system is in the non-degenerate state. 
It has a Maxwell distribution with $\gamma=1.5966$ and the Lorenz number $L_2=1.18 \times 10^{-8}$(W$\cdot\Omega$)/K$^2$ independent of $\vartheta$ again \cite{Redmer1997PR}, and the W-F law is still valid based on the Spitzer formula \cite{Spitzer2006book}. 
In the region of $\vartheta_{lb}\le\vartheta\le\vartheta_{ub}$, however, the system, like WDM, is in the partially degenerate state. 
It has a coexistence of quantum and classical (ions) effects and the W-F law is impossible to be true because the ionized electrons except for conduction electrons emerge and the $L$ varies from $L_1$ descending to $L_2$ with a rise of $\vartheta$. 
We now estimate the approximate boundaries $\vartheta_{lb}$ and $\vartheta_{ub}$ theoretically. 
As mentioned in this text that in the degenerate plasma the electronic thermal de Broglie wavelength $\Lambda_e$ is greater than the averaged electron distance $d_e$, it means $\frac{4\pi}{3}\Lambda_e^3n_e>1$ for the degenerate plasma while $\frac{4\pi}{3}\Lambda_e^3n_e<1$ for the non-degenerate plasma. 
Therefore, $\frac{4\pi}{3}\Lambda_e^3n_e=1$ would be served as a proper condition to estimate approximate boundaries $\vartheta_{lb}$ and $\vartheta_{ub}$. 
As a result, we obtained the upper boundary $\theta_{ub}\sim3.41\bar{z}^{2/3}$ if $n_e$ is the number density of free electrons with $\bar{z}=z^*/z_c$, where $z^*$ is the sum of ionized electrons and conduction electrons $z_c$. 
While the lower boundary is $\theta_{lb}\sim3.41\eta^{-2/3}$ if taking $n_e$ in the normal density and only conduction electrons ($z^*=z_c$). 
The approximation expression of $\vartheta_{lb}$ for pressure ionization shows that the fully degenerate region is contracted while the partially degenerate region is expanded as matter is highly compressed. 
To identify the boundaries $\vartheta_{lb}$ and $\vartheta_{ub}$ is quite important, because one can effectively select the suitable model to exactly describe the nature of WDM.
While the approximate estimation of $\vartheta_{lb}$ and $\vartheta_{ub}$ is possible to fast understanding what kind of matter is WDM without expensive simulations.
 
We now concretely discuss simulation boundaries of warm dense Be at the compressed density $\rho=20$~g/cc, using calculated $\sigma$ and $\kappa$ by the new code EBTP. 
We found from Fig.~\ref{fig: Lorenz number}(d) that the simulated Lorenz number (red stars) completely lies on the $L_1=2.44 \times 10^{-8}$(W$\cdot\Omega$)/K$^2$ (black dished line) as $\vartheta$ varies from 0 to $\sim 0.65$. 
Therefore, the lower boundary of warm dense Be is $\vartheta_{lb} \sim 0.65$ that corresponds to $\alpha \sim 0.063$ in Fig.~\ref{fig: Lorenz number}(c), where calculated $\sigma$ and $\kappa$ are shown in Fig.~\ref{fig: Lorenz number}(a)(b).
The simulated $\vartheta_{lb} \sim 0.65$, very close to the estimated value of 0.69, shows that the fully degenerate region is in $\vartheta\le\vartheta_{lb}\sim 0.65$, where behavior of the compressed Be governed by the conduction electrons indicates that this clearly is not the state of WDM. 
In this region, both $\sigma$ and $\kappa$ are small, as seen in Fig.~\ref{fig: Lorenz number}(a) and \ref{fig: Lorenz number}(b), and the W-F law is available. 
Next, we discuss the upper boundary for warm dense Be. 
With the help of calculated values of $\sigma$ and $\kappa$, the obtained Lorenz number falls on $L_2=1.18 \times 10^{-8}$(W$\cdot\Omega$)/K$^2$ as $\vartheta=\vartheta_{ub}\sim 5.8$, as seen in Fig.~\ref{fig: Lorenz number}(d). 
The upper boundary $\vartheta_{ub}\sim 5.8$ close to the estimated value $\sim 5.2$ corresponds to the degree of ionization $\alpha\sim 0.9$. 
It means that the system for $\vartheta>\vartheta_{ub}\sim 5.8$, is no longer the state of WDM and can be treated by Boltzmann statistics, where the isolated-atom model is suitable and $\sigma$ and $\kappa$ are determined by the Spitzer relationship. 
In this case, $L$ is unchanged with $\vartheta$ and the W-F law is true again. 
Thus, the electrical conductivity is $\kappa=A\sigma T$, and it can be directly obtained from $\sigma$ that is easily measured in experiments, where A is a constant.
We at present discuss properties of warm dense Be in the partially degenerate region of $\vartheta_{lb}\le\vartheta\le\vartheta_{ub}$, or 47.7$<$$T$(eV)$<$407 for $\rho=20$~g/cc, where it must be treated by Fermi-Dirac statistics. 
In such a system, the pressure ionization plays an important role and a great number of ionized electrons emerge with the increase of the degree of ionization of $\alpha$. 
Thus, the thermal flow and electric current with electric field enhance and the electrical conductivity $\sigma$ and thermal conductivity $\kappa$ rise with the increase of $\vartheta$, as seen in Figs.~\ref{fig: Lorenz number}(a-c), resulting in strong transport phenomena in WDM. 
In such a system, the Lorenz number $L$ descends continuously from $L_1$ to $L_2$ till $\vartheta_{ub}\approx 5.8$ and the W-F law is no longer true which leads to the relationship connecting $\kappa$ with $\sigma$ destroyed, completely different from materials that in the full degenerate and non-degenerate regions. 

In summary, we proposed an energy band theory for WDM, which includes the plane wave, the scattering wave, and the degenerated atomic wave, and developed a new code EBTP to investigate properties of WDM. 
The massive data of EOS of WDM in medium and low Z have been obtained. 
We also investigated the transport behavior of WDM using the Lorenz number varying with the degeneracy parameter. 
It is found that the Wiedemann-Franz law for WDM is no longer available due to pressure ionization. 
From the $L$ number we for the first time achieved the upper and lower parameter boundaries of WDM, which are important to identify what kind of matter is WDM and to select suitable models for the study of WDM. 

Acknowledgment: We thank professors Z. M. Sheng and B. Qiao for their fruitful discussions.
This work is financially supported by No.xxx.
We acknowledge the High-performance Computing Platform of Peking University for providing computational resources.

\bibliographystyle{apsrev4-1}
\bibliography{EBT}

\end{document}